\documentclass[aps,prl,onecolumn,superscriptaddress,showpacs]{revtex4}
\usepackage{epsfig,pstricks,dcolumn}
\usepackage{graphicx}
\usepackage{graphics}
\usepackage{graphicx}
\usepackage{epsfig}
\usepackage{epstopdf}

\begin{document}
%\preprint{}

%Title of paper
\title{Quasiparticle light elements and quantum condensates in nuclear matter}
\author{G.~R\"{o}pke}
\email{gerd.roepke@uni-rostock.de}
\affiliation{Institut f\"{u}r Physik, Universit\"{a}t Rostock,
Universit\"{a}tsplatz 3, D-18055 Rostock, Germany}

\begin{abstract}
Nuclei in dense matter are influenced by the medium. 
In the cluster mean field approximation, 
an  effective Schr\"odinger equation for the $A$-particle cluster is obtained 
accounting for the effects of the surrounding medium, such as
self-energy and  Pauli blocking. Similar to the single-baryon states 
(free neutrons and protons), the light elements ($2 \le A \le 4$, internal quantum state $\nu$)
are treated as quasiparticles with energies  $E_{A,\nu}(P; T, n_n,n_p)$ 
that depend on the center of mass momentum $\vec P$, the temperature $T$, 
and the total densities $n_n,n_p$ of neutrons and protons, respectively. 
We consider the composition and thermodynamic properties of nuclear matter 
at low densities. At low temperatures, quartetting is expected to occur. 
Consequences for different physical properties of nuclear matter and finite nuclei are discussed.
\end{abstract}
\date{\today}

\maketitle

 \section{\label{sec:introduction}
 Introduction}

Quantum condensates are one of the amazing phenomena in many-particle physics.
Isospin-triplet (proton-proton or neutron-neutron) pairing is well established \cite{RS} 
not only in nuclear matter at low temperatures, but also in finite nuclei. Isospin-singlet (neutron-proton) pairing
may become of relevance in symmetric and asymmetric nuclear matter \cite{triplet}. On the one hand, the interaction is stronger, 
even a bound state, the deuteron, can be formed. This would lead to higher transition temperatures and the transition from 
Cooper pairing to Bose-Einstein condensation. On the other hand, with increasing difference of the chemical potentials and increasing Coulomb effects,
the formation of a condensate becomes  more difficult, in particular in asymmetric matter. The strong interaction in the isospin-singlet channel
leads to a precursor for the Bose condensate, the occurrence of a pseudogap \cite{pseudo} above the critical temperature.
Furthermore,  pairing competes with the formation of an $\alpha$-particle condensate (quartetting) \cite{RSSN}, in particular at low densities.
Therefore, the role of correlations, in particular the formation of clusters and of a quantum condensate, is an interesting item in
strongly interacting quantum liquids such as nuclear matter. 

Systematic quantum statistical approaches, see \cite{R}, have to be used for a 
treatment using appropriate concepts such as thermodynamic Green functions, spectral function, and self-energy. Within a chemical picture,
a cluster decomposition of the self-energy can be performed that considers the formation of clusters in a dense medium, taking into account 
symmetrization (Pauli blocking), screening of the interaction, dynamical self-energy, and other many-particle effects. An interesting 
approximation is the cluster mean-field approximation \cite{R,cmf,DRS} that considers the few-particle Schr\"odinger equation in a correlated medium, 
where the correlations in the medium have to be determined in a self-consistent way. This would allow to describe nuclear matter in a wide range of 
densities, from saturation density where a mean-field approach on the single nucleon level is possible, to the low-density region where 
the nuclear statistical equilibrium (NSE) applies and $\alpha$-matter 
can occur. We discuss the formation of clusters in nuclear matter in Sec. 2 and give some applications to finite nuclei in Sec. 3.

\section{Nuclei in matter}
\label{sec: cluster}

The few-body problem describing $A \le 4$ nucleons in hot and dense matter can be related to an in-medium wave equation
(Bethe-Salpeter equation) that is derived from many-particle approaches, see Ref.~\cite{R}.
We consider only bound states $\nu$, dropping the spin quantum number.
For the light elements, $\nu= d, t, h, \alpha$ denotes the deuteron ($^2$H), the triton ($^3$H), the helion ($^3$He), and the $\alpha$ particle  ($^4$He). 
Considering uncorrelated nucleons in the medium, the few-nucleon wave function 
and the corresponding eigenvalues follow from solving the in-medium
Schr\"odinger equation 
\begin{eqnarray}
&&[E_1^{\rm qp}(1)+\dots + E_1^{\rm qp}(A)]\psi_{\nu P}(1\dots A)
\nonumber \\ &&
+\sum_{1'\dots A'}\sum_{i<j}[1- f_1(i)-  f_1(j)]V(ij,i'j')\prod_{k \neq 
  i,j} \delta_{kk'}\psi_{ \nu P}(1'\dots A')
%\nonumber \\ &&
= E^{\rm qp}_{\nu}(P) \psi_{ \nu P}(1\dots A)\,.
\label{waveA}
\end{eqnarray}
For  brevity, the single-nucleon quasiparticle energy $ E_{\tau_1}^{\rm qp}(\vec p_1) $ is denoted as $ E_1^{\rm qp}(1) $.
The nucleon-nucleon interaction $V(ij,i'j')$ becomes medium dependent due to the Pauli blocking prefactor 
$[1- f_1(i)- f_1(j)]$.
The phase space occupation is described by a Fermi distribution function normalized to the total density of nucleons,
\begin{equation}
\label{f1}
 f_1(1) = \frac{1}{\exp[E_1^{\rm
qu}(1)/T-  \mu_\tau/T] +1} \approx \frac{n_\tau}{2} \left(\frac{2 \pi \hbar^2}{m_\tau T}\right)^{3/2} 
e^{-\frac{ p_1^2}{2 m_\tau T}}
\end{equation}
in the low-density, non-degenerate limit ($ \mu_\tau <0  $). The chemical potential  $ \mu_\tau$ is
determined by the normalization condition $2 \Omega^{-1} \sum_p f_1(p) =
n_{\tau}$, where $\tau $ denotes isospin (neutron or proton), and has to be expressed in terms of these densities $n_n, n_p$ 
or the baryon density $n=n_n+n_p$ and proton fraction $Y_p=n_p/n$, and the temperature $T$.

The  in-medium Schr\"odinger equation (\ref{waveA}) contains the effects of the medium in the single nucleon quasiparticle shift 
as well as in the Pauli blocking terms. 
Obviously, the bound state wave functions and energy eigenvalues as
well as the scattering phase shifts depend on temperature
and density. In particular, we obtain the cluster quasiparticle shifts 
\begin{equation}
\label{quasi}
E_\nu^{\rm qp}(P;T,n,Y_p)-E_\nu(P)
=\Delta E_\nu^{\rm SE}(P;T,n,Y_p)+\Delta E_\nu^{\rm Pauli}(P;T,n,Y_p). 
\end{equation}

The contribution of the single nucleon  energy shift to the cluster self-energy shift $\Delta E_\nu^{\rm SE}$ 
is easily calculated in the effective mass approximation, 
where the single-nucleon quasiparticle energy shift $\Delta E_1^{\rm SE}(1)$ 
can be represented by the energy shift 
$\Delta E_\tau^{\rm SE}(T,n,Y_p)$ and the effective mass $m^*_\tau(T,n,Y_p)$, see \cite{Rarxiv}.  
Different expressions for the single-nucleon quasiparticle shifts are available such as Skyrme, 
RMF, or DBHF, see \cite{Typel}. Since the  single-nucleon quasiparticle energy shift arises in the 
continuum of scattering states as well, the influence on the binding energies is small.

We consider here the Pauli blocking shift of the binding energies 
\begin{equation}
\label{Paul}
\Delta E_\nu^{\rm Pauli}(P;T,n,Y_p)=E_\nu^{\rm qp}(P)-E_\nu(P)- (A-Z) \Delta E_n^{\rm SE}- Z \Delta E_p^{\rm SE}\,.
\end{equation}
Different fit formula have been given \cite{R,Typel,SR} to parametrize the solution of the in-medium Schr\"odinger equation (\ref{waveA}).
We follow the recent work \cite{Rarxiv}. 

In the low-density limit, a linear dependence of the energy shifts 
on the nucleon density $n$ follows from  perturbation theory. For a more general dependence on the total nucleon density
we consider the expression
\begin{eqnarray}
\label{delpauli0P2}
&&\Delta E_\nu^{\rm Pauli}(P;n,T,Y_p)= 
c_{\nu}(P;T) \left\{1-\exp\left[- \frac{f_\nu(P;T,n)}{c_{\nu}(P;T)} y_\nu(Y_p)n- d_{\nu}(P;T,n)n^2 \right] \right\}\,,
\end{eqnarray}
where the dependence on the asymmetry is given by $y_d(Y_p)=y_\alpha(Y_p)=1$, for triton $y_t(Y_p)=\left( \frac{4}{3}-\frac{2}{3} Y_p\right)$, 
and for helion $y_h(Y_p)=\left(\frac{2}{3}+\frac{2}{3} Y_p\right)$ that reflects the different proton and neutron content..
The functions $f_\nu(P;T,0)$ can be calculated in first order perturbation theory using the unperturbed wave 
functions of the free nuclei. We performed model calculations with a separable potential  that reproduces the empirical binding energies and
rms radii. Motivated by the 
exact solution for $A=2$, we use the following fit for arbitrary $\nu$:
\begin{eqnarray}
\label{perturb}
f_\nu(P;T,n)&=& f_{\nu,1}  \exp\left[-\frac{P^2/\hbar^2}{4 ( f_{\nu,4}^2/f_{\nu,3}^2) (1+T/f_{\nu,2})+u_\nu n}\right] \frac{1}{T^{1/2}} \frac{2 f_{\nu,4}}{P/\hbar}   
 {\rm Im} \left\{e^{x^2}  {\rm erfc}\left[ x\right]\right\}\,,
 \nonumber\\
x&=& f_{\nu,3} (1+f_{\nu,2}/T)^{1/2} \left(1-i\frac{P/\hbar}{2 f_{\nu,4}(1+T/f_{\nu,2})}\right)\,.
\end{eqnarray}
%%%%%%%%%%%%%%%%%%%%%%%%%%%%%%%%%%%%%%%%%%%%%%%%%%%%%%%%%%%%%%%
\begin{table}
\caption{ Parameter values for the Pauli blocking shift $\Delta E^{\rm Pauli}_\nu(P;T,n,Y_p)$, Eq.~(\ref{perturb}), in the low-density limit}
\begin{center}
\hspace{0.5cm}
\begin{tabular}{|c|c|c|c|c|c|}
\hline
 $\nu$  & $f_{\nu,0}$  &$ f_{\nu,1}$  & $f_{\nu,2}$ & $f_{\nu,3}$ & $f_{\nu,4}$ \\
 & [MeV fm$^{5/2}$]& [MeV fm$^3$] & [MeV] & -& [fm$^{-1}]$\\
\hline
$d$ ($^2$H) & 388338  & 6792.6 &  22.52 &0.2223 &0.2317\\ 
 $t$ ($^3$H) &  159080 &   20103.4 & 11.987 &0.85465  &0.9772\\
 $h$ ($^3$He) &153051 &19505.9 &  11.748  & 0.84473  &0.9566 \\
$\alpha$ ($^4$He)  & 352965 & 36146.7 & 17.074  &0.9865&1.9021\\
\hline
\end{tabular}
\label{Tab.2neu}
\end{center}
\end{table}

%%%%%%%%%%%%%%%%%%%%%%%%%%%%%%%%%%%%%%%%%%%%%%%%%%%%%%%%%%%%%%%
For zero momenta, $P=0$, the temperature dependence of $c_{\nu}(0;T)$ and $d_{\nu}(0;T,n)$ is expressed as
\begin{equation}
c_{\nu}(0;T)=c_{\nu,0}+\frac{c_{\nu,1}}{(T-c_{\nu,2})^2+c_{\nu,3}},\qquad d_{\nu}(0;T,n)=\frac{d_{\nu,1}}{(T-d_{\nu,2})^2+d_{\nu,3}}\,.
\end{equation}
The Pauli shift at finite momenta is fitted with  $c_{\nu}(P;T)=c_{\nu}(0;T)$ not depending on $P$, but
\begin{equation}
 d_{\nu}(P;T,n)=d_{\nu}(0;T,n)e^{-\frac{P^2/\hbar^2}{v_\nu T n}}\,.
\end{equation}
An additional dependence on $n,T$ is considered at finite values of $P$. 
Another additional dependence on $n,T$ for finite momenta is introduced in $f_\nu(P;T,n)$ where the dispersion relation becomes density dependent due to the parameter $u_\nu$.
Parameter values are given in Tab.~\ref{Tab.2neu} and Tab.~\ref{Tabcd}. 

\begin{table}
\caption{ Parameter values for the Pauli blocking shift $\Delta E^{\rm Pauli}_\nu(0;T,n,Y_p)$, Eq.~(\ref{delpauli0P2}),
in units of MeV and fm ($c_{\nu,0},c_{\nu,2},d_{\nu,2}$ - [MeV];
$c_{\nu,1}$ - [MeV$^3$]; $c_{\nu,3},d_{\nu,3}$ - [MeV$^2$]; $d_{\nu,1}$ - [MeV$^2$ fm$^6$]; $u_\nu$ - fm; $v_\nu$ - [MeV$^{-1}$ fm])}
\begin{center}
\hspace{0.5cm}
\begin{tabular}{|c|c|c|c|c|}
\hline
 $\nu$  &$d$ ($^2$H) & $t$ ($^3$H) & $h$ ($^3$He)&$\alpha$ ($^4$He) \\
\hline
 $c_{\nu,0}$ &2.752 &11.556&10.435 &150.71 \\
 $c_{\nu,1}$ &32.032  & 117.24 &176.78  &9772 \\
 $c_{\nu,2}$   & 0 & 3.7362 &  3.5926 &2.0495 \\
$c_{\nu,3}$ & 9.733 & 4.8426 & 5.8137  &2.1624\\
\hline
$d_{\nu,1}$  & 523757  & 108762 & 90996  &5391.2 \\
 $d_{\nu,2}$  &0&  9.3312 &  10.72 &    3.5099 \\
 $d_{\nu,3}$ & 15.273 &49.678 &47.919 & 44.126\\
\hline
$u_{\nu}$  &11.23  &25.27 & 25.27  &44.92 \\
\hline
$v_{\nu}$  &0.145  &0.284 & 0.27  &0.433 \\
\hline
\end{tabular}
\label{Tabcd}
\end{center}
\end{table}

The in-medium Schr\"odinger equations (\ref{waveA}) describe in the case $ A=2 $ ($d$) and  $ A=4 $ ($\alpha$)
also the onset of a quantum condensate if according to the Thouless condition the in-medium binding energy
coincides with $ A \mu $. 

With these parameter values, it is possible to evaluate the composition of warm nuclear matter at arbitrary baryon density
$n $ below the saturation density, temperature $T$, and proton fraction $Y_p$. Equations of state can be obtained for 
different thermodynamic properties, see also \cite{Typel}. A smooth transition from the description of nuclear matter 
near saturation density to the low density region where the NSE can be applied has been found. A systematic inclusion of mean-field effects is possible.

An unsolved problem is the behavior of the internal energy at low temperatures.
The internal energy for symmetric matter is shown in Fig. 1. At low densities and temperatures it approaches the binding energy of the $\alpha$ particle.
With increasing density, the self-energy shift leads to the behavior $\partial U/\partial n <0$ at $T=0$ that means thermodynamical instability. With the disappearance of the bound states, the internal energy approaches the values for the quasiparticle single-nucleon approximation. This occurs rather 
abruptly near $n= 0.02$ fm$^{-3}$ so that an intermediate region of metastability appears. However, this may be changed if better approximations for the 
internal energy in the intermediate region are available.  Alternative approaches to  the internal energy of symmetric matter at $T=0$ have been considered in 
connection with the formation of a quantum condensate \cite{Toh,Mis}.

\begin{figure}
\begin{center}
\includegraphics[width=10cm]{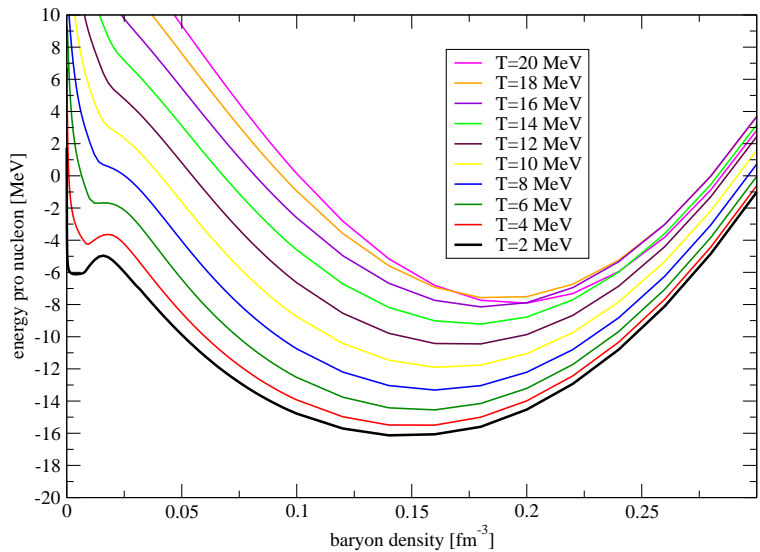}% Here is how to import EPS art
\caption{\label{fig:1} 
Internal energy $U$ for symmetric nuclear matter at different temperatures.}
\end{center}
\end{figure}

\section{$\alpha$ clustering in nuclei}

Nuclear matter at low densities and temperatures is thermodynamically instable. This region of phase instability can only be reached
in nonequilibrium or in inhomogeneous systems. An interesting application are nuclei, where nuclear systems at low densities can occur. 
In particular, excited states
of symmetric 4$n$ nuclei near the $n \alpha $ threshold may show an $\alpha$ cluster state such as in $^8$Be, $^{12}$C, possibly $^{16}$O, and others.
As a simple ansatz to describe $\alpha $ clusters with the same c.o.m. orbits, like the pairing state, the THSR wave function was introduced \cite{THSR}.
This wave function is a product ansatz for the c.o.m. motion of antisymmetrized $\alpha$ clusters, similar to the BCS wave function.
This approach has been proven to describe different properties of such low-density states successfully.

In particular, a relation between the density and the condensate fraction has been discussed \cite{cond} that indicates the disappearence of the condensate
around $n_{\rm saturation}/3$. Whereas $\alpha$ clusters are well developed in the Hoyle state, in the ground state of $^{12}$C no $\alpha$ clustering
is present and a quasiparticle single-nucleon shell model is applicable. The reason is the antisymmetrization of the nucleonic wave functions
that results in the Pauli blocking. In contrast to the calculations shown in Fig. 1, where the Pauli blocking is caused by uncorrelated nucleons that
form a Fermi sphere, the Pauli blocking by a clustered medium is less efficient because the nucleonic wave function is more distributed in momentum space.
Therefore, we expect that the crossover from $\alpha$ matter behavior, given by a cluster-virial expansion, to the quasiparticle nucleonic liquids
happens not near $\rho = 0.02$ fm$^{-3}$, but near  $\rho = 0.05$ fm$^{-3}$ so that a continuous decrease is expected. Then, the intermediate region of
metastability where $\partial \mu/\partial n \ge 0$ would disappear. 

The behavior of the quasiparticle shift $E_\alpha^{\rm qp}(P;T,n,Y_p)-E_\alpha(P)  $, Eq. (\ref {quasi}), is sensitive to the values of self-energy shift and Pauli blocking that partially compensate. At zero temperature, the self-energy shift amounts about -4000~$n$  MeV fm$^3$, whereas the Pauli blocking term 
is larger than 4434~$n$   MeV fm$^3$ as given for $T=1$ MeV in \cite{R}. If we estimate the kinetic energy of the nucleons bound in the $\alpha$ particle 
as $E_\alpha^{\rm kin}$ = 51. 6 MeV \cite{Rarxiv} and simulate the corresponding occupation in momentum space by 
an effective temperature $E_\alpha^{\rm kin}/6$, the Pauli blocking shift is reduced by a factor 2 so that the total quasiparticle energy shift becomes 
negative, around -2000~$n$ MeV fm$^3$ at $T=0$. A decrease of the internal energy follows also from the cluster virial expansion \cite{HS} that gives about
-1500~$n$  MeV fm$^3$ at $T=1$ MeV. This value is only based on empirical $\alpha-\alpha$ scattering phase shits. 
The Ali-Bodmer interaction, which is also based on these phase shifts, has been considered in \cite{Mis} to calculate the $\alpha$ condensate fraction at $T=0$.
Also a decrease of internal energy was found. Exploratory lattice calculations for a $T=0$ $\alpha$ condensate state \cite{Toh}  gave no definite  answer.
Whereas for $^4$He an increase of energy with increasing density was obtained, the opposite occurs for $^{16}$O. 

Another interesting region where $\alpha$ clustering is expected to occur is the surface of nuclei. Like pairing that is increasing near the surface \cite{RS} 
where the density drops down from the saturation value, also $\alpha$ clustering  and possibly quartetting can occur. A simple way to incorporate this is 
a local density approach. Estimations of the role of isospin-singlet pairing and quartetting for nuclei near the $N=Z$ line have been given in \cite{LWigner} 
analysing binding energies of the ground state and of the excited state. The Wigner energy can be explained this way. This effect is seen in the light nuclei
where the shift of the proton chemical potential due to the Coulomb energy is not strong compared to the condensation energy.

Cluster formation at the surface of nuclei is of interest also in the region of superheavy elements. As well known, the most dominant decay channel is
$\alpha$ decay, in contrast to the emission of protons or neutrons. Therefore one expects that $\alpha$ particles are preformed in the skin of the superheavies
where the density drops down. A corresponding picture is a stable core such as the double magic lead or uranium nucleus, surrounded by a skin of nucleons
at reduced density where cluster formation is possible. Of course, the Pauli blocking mechanism has to be considered beyond the local density approximation,
considering the density matrix due to the core orbitals. We discuss two consequences of clustering in the skin of superheavies, the binding energy and the rms 
radii. Further signatures can be seen in reaction processes.

The staggering of the binding energies of superheavies is well known from experimental data. It would be of interest to what extent this staggering can be 
explained within a quasiparticle single-nucleon approach such as shell calculations, or using a THSR-like wave function that describes the c.o.m. motion of
antisymmetrized clusters. 

Another effect are the rms radii of superheavy elements. Experimentally, the charge radii $\langle r^2 \rangle_{\rm charge}$ are measured. The calculated
values are the point radii $\langle r^2 \rangle_{\rm point}$ that consider point-like nucleons. The relation between both is given by the relation \cite{Angeli}
\begin{equation}
\langle r^2 \rangle_{\rm point}+\langle r^2 \rangle_{\rm proton}+(N/Z)\langle r^2 \rangle_{\rm neutron} =\langle r^2 \rangle_{\rm charge}
\end{equation}
that takes into account the contribution of the proton form factor due to its spatial extension, $\langle r^2 \rangle_{\rm proton}=0.743$ fm$^2$, 
and the neutron contribution $\langle r^2 \rangle_{\rm neutron}=-0.116$ fm$^2$. If we assume that the surface is formed by $\alpha$-like clusters,
we expect another relation,
\begin{equation}
\langle r^2 \rangle_{\rm point}+\langle r^2 \rangle_{\rm \alpha} =\langle r^2 \rangle_{\rm charge}
\end{equation}
with the charge rms radius for the $\alpha $ particle, $\langle r^2 \rangle_{\rm \alpha} = 2.822$ fm$^2$. 
Assuming that the point rms radii are nearly the same, the effect of $\alpha$-clustering would increase the rms radius according
$\langle r^2 \rangle_{\rm \alpha}-\langle r^2 \rangle_{\rm proton}-(N/Z)\langle r^2 \rangle_{\rm neutron}=2.195$ fm$^2$.
For instance, considering a heavy nucleus with $[\langle r^2 \rangle_{\rm point}]^{1/2}=6$ fm, the charge rms radius would be larger
by 2.98 \% if the $\alpha$ cluster determine the surface, compared to the usual nucleonic surface. The empirical values of charge rms radii
of nuclei \cite{Angeli} have been compared with Hartree-Fock Bogoliubov shell model calculations \cite{HFB}, and excellent agreement has been found with 
exception of the superheavies like Cm where 
the measured values of the charge rms radii are about 2 \% higher than the calculated ones. An interesting point would be whether the 
difference can be explained by clustering. Assuming this, $\alpha$ emitters should have an expanded charge rms radius, compared to 
quasiparticle single-nucleon calculations.\\

%\noindent {\bf References}\\

%%%%%%%%%%%%%%%%%%%%%%%%%%%%%%%%%%%%%%%%%

\end{document}